\documentclass[conference]{IEEEtran}
\IEEEoverridecommandlockouts
\ifCLASSINFOpdf
\else
\fi
\usepackage[top=0.75in, bottom=0.75in, left=0.75in, right=0.75in]{geometry}
\usepackage{bbm}
\usepackage{verbatim}
\usepackage{graphicx}
\usepackage{cite}
\usepackage{url}
\usepackage[cmex10]{amsmath}
\usepackage{amssymb}
\usepackage{amsmath}
\usepackage{amsthm}
\usepackage{algorithm, algorithmicx, algpseudocode}
\usepackage[caption=false,font=footnotesize]{subfig}
\usepackage{color}
\usepackage{cite}
\usepackage{epstopdf}
\usepackage{calc}
\usepackage{array}
\usepackage{stfloats}
\usepackage{siunitx}
\usepackage{mathtools}
\usepackage{optidef}
\usepackage{gensymb}
\usepackage[printonlyused]{acronym}
\usepackage{xcolor}
\usepackage{cleveref}
\usepackage{multirow}

\acrodef{mmW}{Millimeter-wave}
\acrodef{BS}{base station}
\acrodef{UE}{user equipment}
\acrodef{SOTA}{state of the art}
\acrodef{AoA}{angle of arrival}
\acrodef{AoD}{angle of departure}
\acrodef{AWV}{antenna weight vector}
\acrodef{ADC}{analog-to-digital converter}
\acrodef{BB}{baseband}
\acrodef{RSRP}{reference signal received power}
\acrodef{CSI}{channel state information}
\acrodef{COTS}{commercial-off-the-shelf}
\acrodef{PAA}{phased antenna array}
\acrodef{TTD}{true-time-delay}
\acrodef{LoS}{line-of-sight}
\acrodef{NLoS}{non-line-of-sight}
\acrodef{IA}{initial access}
\acrodef{DFT}{discrete Fourier transform}
\acrodef{UDN}{ultra-dense networks}
\acrodef{RF}{radio frequency}
\acrodef{MPC}{multipath component}
\acrodef{BF}{beamforming}
\acrodef{SNR}{signal-to-noise ratio}
\acrodef{SINR}{signal-to-interference-plus-noise ratio}
\acrodef{OFDM}{orthogonal frequency-division multiplexing}
\acrodef{DSP}{digital signal processing}
\acrodef{LUT}{lookup table}
\acrodef{MIMO}{multiple-input multiple-output}
\acrodef{IC}{integrated circuits}
\acrodef{PS}{phase shifter}
\acrodef{DAC}{digital-to-analog converter}
\acrodef{EVM}{error vector magnitude}
\acrodef{CP}{cyclic prefix}
\acrodef{FPGA}{field programmable gate arrays}
\acrodef{MSE}{mean squared error}
\acrodef{RMSE}{root mean square error}
\acrodef{MMSE}{minimum mean square error}
\acrodef{VTC}{voltage-to-time converter}
\acrodef{TDC}{time-to-digital converter}
\acrodef{CMOS}{complementary metal–oxide–semiconductor}
\acrodef{HI}{hardware impairments}
\acrodef{CS}{sampling capacitor}
\acrodef{RST}{reset}
\acrodef{SCA}{Switched-Capacitor Arrays}

\DeclareMathOperator*{\argmin}{argmin}
\DeclareMathOperator*{\argmax}{argmax}

\DeclarePairedDelimiter{\ceil}{\lceil}{\rceil}
\DeclarePairedDelimiter{\floor}{\lfloor}{\rfloor}
\DeclarePairedDelimiter\abs{\lvert}{\rvert}

\newcommand\norm[1]{\left\lVert#1\right\rVert}

\newcommand{\BW}[0]{\mathrm{BW}}
\newcommand{\BWc}[0]{\mathrm{BW}_{\text{c}}}
\newcommand{\sigmaN}[0]{\sigma_{\text{N}}}
\newcommand{\SNR}[0]{\mathrm{SNR}}
\newcommand{\fc}[0]{f_{\text{c}}}
\newcommand{\T}[0]{\text{T}}
\newcommand{\R}[0]{\text{R}}
\newcommand{\I}[0]{\text{I}}
\newcommand{\hermitian}[0]{\text{H}}
\newcommand{\transpose}[0]{\text{T}}
\newcommand{\tot}[0]{\text{tot}}

\begin{document}
%

\title{Design of Millimeter-Wave Single-Shot Beam Training for True-Time-Delay Array}


\author{
\IEEEauthorblockN{Veljko Boljanovic\IEEEauthorrefmark{1}, Han Yan\IEEEauthorrefmark{1}, Erfan Ghaderi\IEEEauthorrefmark{2}, Deukhyoun Heo\IEEEauthorrefmark{2}, Subhanshu Gupta\IEEEauthorrefmark{2}, and Danijela Cabric\IEEEauthorrefmark{1}}
\IEEEauthorblockA{\IEEEauthorrefmark{1}Electrical and Computer Engineering Department, University of California, Los Angeles, CA, USA}
\IEEEauthorblockA{\IEEEauthorrefmark{2}School of Electrical Engineering and Computer Science, Washington State University, WA, USA\\
Email: \{vboljanovic, yhaddint\}@ucla.edu, \{erfan.ghaderi, dheo\}@wsu.edu, sgupta@eecs.wsu.edu, danijela@ee.ucla.edu}

}



\IEEEoverridecommandlockouts

\maketitle

\begin{abstract}
Beam training is one of the most important and challenging tasks in millimeter-wave and sub-terahertz communications. Novel transceiver architectures and signal processing techniques are required to avoid prohibitive training overhead when large antenna arrays with narrow beams are used. In this work, we leverage recent developments in wide range true-time-delay (TTD) analog arrays and frequency dependent probing beams to accelerate beam training. We propose an algorithm that achieves high-accuracy angle of arrival estimation with a single training symbol. 
Further, the impact of TTD front-end impairments on beam training accuracy is investigated, including the impact of gain, phase, and delay errors. Lastly, the study on impairments and required specifications of resolution and range of analog delay taps are used to provide a design insight of energy efficient TTD array, which employs a novel architecture with discrete-time sampling based TTD elements.

\end{abstract}


%
\IEEEpeerreviewmaketitle
%
%
\section{Introduction}
\label{sec:introduction}
\ac{mmW} communications have the key role in providing high data rates in the fifth generation of cellular systems due to abundant spectrum. MmW systems require beamforming with large antenna arrays at both the \ac{BS} and \ac{UE} to combat severe propagation loss. The information of the \ac{AoD} and \ac{AoA} is necessary to enable directional communication between the \ac{BS} and \ac{UE}. Such information is typically acquired by beam training in standardized \ac{mmW} systems. However, with increased array size and reduced beam width, the training overhead increases. The challenge of overhead will become even more severe in the future \ac{mmW} and sub-terahertz systems where the array size is expected to further increase \cite{6G_intro_NYU}.

The \ac{mmW} beam training is an active research area. Various algorithms have been developed for phase shifter based analog arrays to reduce the training overhead to logarithmically scale with the number of antenna elements. Some of the proposed algorithms include iterative search \cite{7845674} and compressive sensing based search \cite{Yan_CSIA}.
Recent work aimed to reduce the overhead to as few as one training symbol using novel transceiver architecture and algorithm design. Examples include fully digital arrays with low resolution data converter \cite{NYU_IA_digital_array}, and leaky wave antennas \cite{leaky_wave_THz}. Although digital array is an appealing architecture for the \ac{BS} \cite{Yan_digital_array}, it may not suitable for the \ac{UE}. The leaky wave antenna can only be used for beam training and other dedicated circuits are required in the data communication. In our previous work \cite{Yan:TTD}, we showed that by replacing phase shifters with true-time delay elements, the analog array can also implement single-shot beam training. 
In this work, we further investigate \ac{TTD} based single-shot beam training. This work has two major contributions. Firstly, we developed a super-resolution algorithm that improves the \ac{RMSE} of \ac{AoA} estimates as compared to \cite{Yan:TTD}.
The proposed algorithm is also more robust to frequency selective fading in wideband channels. 
Secondly, we study the impact of \ac{TTD} \ac{HI}, including the gain and phase mismatch and \ac{TTD} control error. We also discuss the implementation of TTD arrays and feasibility of hardware specifications for wideband mmW operation.



The rest of the paper is organized as follows. In \Cref{sec:models}, we present the system model of wideband beam training using \ac{TTD} based array architecture. In \Cref{sec:awv_design}, we explain how \ac{AWV}s with frequency diversity are designed. The proposed high-resolution beam training algorithm is presented in \Cref{sec:algorithm}. The performance results are presented in \Cref{sec:results}. In \Cref{sec:implementation}, the \ac{TTD} hardware implementation details are discussed. Finally, \Cref{sec:conclusions} concludes the paper.


Scalars, vectors, and matrices are denoted by non-bold, bold lower-case, and bold upper-case letters, respectively.
The $(i,j)$-th element of $\mathbf{A}$ is denoted by $[\mathbf{A}]_{i,j}$. Conjugate, transpose, Hermitian transpose are denoted by $(.)^{*}$, $(.)^{\transpose}$, and $(.)^{\hermitian}$, respectively.

%
%
\section{System Model}
\label{sec:models}

We consider downlink beam training between a \ac{BS} and a \ac{UE}, where the \ac{CP} based \ac{OFDM} waveform is used. The carrier frequency, bandwidth, and number of subcarriers are denoted as $\fc$, $\BW$, and $M_{\tot}$, respectively. Both the \ac{BS} and \ac{UE} have half-wavelength spaced uniform linear arrays with $N_{\T}$ and $N_{\R}$ antennas, respectively.

\begin{figure}
    \begin{center}
        \includegraphics[width=0.49\textwidth]{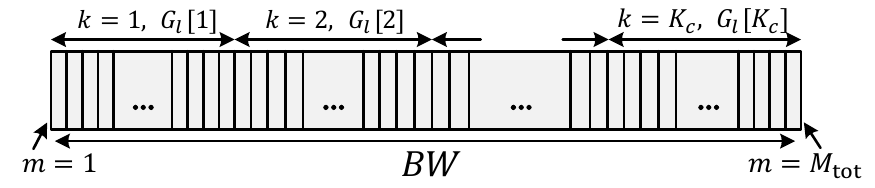}
    \end{center}
    \vspace{-4mm}
    \caption{The illustration of sub-bands and their corresponding channel gains for the $l$-th cluster.}
    \vspace{-4mm}
    \label{fig:indices}
\end{figure}

We consider a frequency selective geometric channel model with $L$ multipath clusters. Assuming coherence bandwidth $\BWc$, $\BW$ can be segmented into $K_c=\ceil*{\BW/\BWc}$ distinct sub-bands with different channels. We assume that all OFDM subcarriers within the $k$-th sub-band experience the same channel $\mathbf{H}[k]\in\mathbb{C}^{N_{\R}\times N_{\T}}$, which can be expressed as
\begin{equation}
\label{eq:channel1}
    \mathbf{H}[k] = \sum_{l=1}^L G_l[k] \mathbf{a}_{\R}(\theta_l^{(\R)})\mathbf{a}_{\T}^{\hermitian}(\theta_l^{(\T)}),
\end{equation}
where $\theta_l^{(\R)}$ and $\theta_l^{(\T)}$ are the \ac{AoA} and \ac{AoD} of the $l$-th cluster. According to illustration in \Cref{fig:indices}, the relationship between the sub-band index $k$ and subcarrier index $m$ is given as $k=\ceil*{(mK_c)/M_{\tot}}$. In this work, we assume the array responses are frequency flat, i.e., $[\mathbf{a}_{\R}(\theta)]_n = N^{-1/2}_{\R} \mathrm{exp}({-j(n-1)\pi\sin({\theta)}}),~n=1,...,N_{\R}$ and $[\mathbf{a}_{\T}(\theta)]_n = N^{-1/2}_{\T}\mathrm{exp}({-j(n-1)\pi\sin{(\theta)}}),~n=1,...,N_{\T}$. 
The complex gains $G_l[k],~\forall k$, come from the multipath rays within the $l$-th cluster. For tractable algorithm design, we assume that $G_l[k]\sim\mathcal{CN}\left(0, \sigma^2_l \right),~\forall k$, are uncorrelated elements of a $K_c$-dimensional multivariate complex Gaussian vector. Further, we assume the complex gains are independent across different clusters, so the covariance between channel gains can be expressed as
\begin{align}
\label{eq:covariance}
    \mathbb{E}\left(G_{l_1}[k_1]G^{*}_{l_2}[k_2]\right) = 
    \begin{cases}
    \sigma^2_{l_1}, & \text{if } l_1=l_2, k_1=k_2\\
    0, & \text{otherwise.} 
    \end{cases}
\end{align}
For the rest of this paper, we assume that the clusters are ordered from the strongest to the weakest one, i.e., $\sigma_1^2 \geq \cdots \geq \sigma_L^2$, without loss of generality.

\subsection{Received signal model with TTD array}
\label{models:signal_impairments}
In this work, we focus on the receiver beam training by assuming \ac{AoD} estimate $\hat{\theta}^{(\T)}$ is available to design a phased array based analog precoder $\mathbf{v} = \mathbf{a}_{\T}(\hat{\theta}^{(\T)})$ at the \ac{BS}. The BS utilizes the same non-zero power-normalized training pilot at all $M$ subcarriers from the predefined set $\mathcal{M}$, i.e., $X[m]=M^{-1/2},~m\in\mathcal{M}$, where $M = |\mathcal{M}|$.

\begin{figure}
    \begin{center}
        \includegraphics[width=0.48\textwidth]{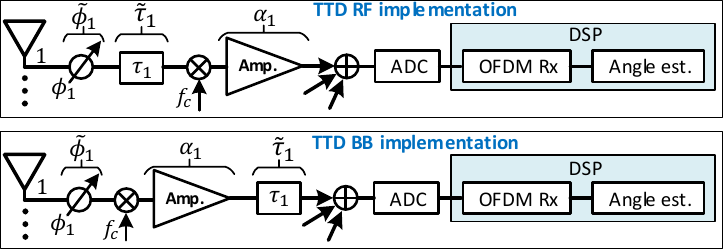}
    \end{center}
    \vspace{-4mm}
    \caption{Two architecture candidates for TTD receiver implementation and the corresponding hardware impairments.}
    \vspace{-4mm}
    \label{fig:architecture}
\end{figure}

The \ac{UE} is equipped with an analog \ac{TTD} array and it performs beam training to estimate the \ac{AoA} $\hat{\theta}^{(\R)}$. As illustrated in \Cref{fig:architecture}, each receiver branch has a phase shifter, local oscillator, set of amplifiers, and \ac{TTD} element. The \ac{TTD} module introduces a group delay $\tau_n$, and the phase shifter introduces a phase shift $\phi_n$. Our previous work \cite{Yan:TTD} showed that when the \ac{CP} is longer than the cumulative delay from channel multipath and TTD elements, the received signal $Y[m]$ at the $m$-th subcarrier is
\begin{equation}
    Y[m] = M^{-1/2}\mathbf{w}^{\hermitian}[m] \mathbf{H}[k] \mathbf{v} + \mathbf{w}^{\hermitian}[m]\mathbf{n}[m],~ m\in \mathcal{M}.
\label{eq:received_signal_sc}
\end{equation}
where $k$ is a sub-band index, as discussed earlier. The thermal noise at the $m$-th subcarrier is denoted as $\mathbf{n}[m]\sim \mathcal{CN}(0,\sigma^2_{\text{N}}\mathbf{I}_{N_{\R}})$. The vector $\mathbf{w}[m]\in\mathbb{C}^{N_{\R}}$ is a \ac{TTD} frequency dependent \ac{AWV}, whose $n$-th element is expressed as \cite{Yan:TTD}
\begin{align}
    [\mathbf{w}[m]]_n 
    = \mathrm{exp}\left[-j (2\pi f_m\tau_n + \phi_n\right)],
    \label{eq:TTD_AWV_wo_error}
\end{align}
where $f_m =\fc - \BW/2 + (m-1)\BW/(M_{\tot}-1)$.

Furthermore, we are interested in understanding the impact of TTD \ac{HI} on the beam training. As illustrated in \Cref{fig:architecture}, we consider two TTD array architectures: \ac{RF} \ac{TTD} array where the group delay is introduced in the \ac{RF} domain and \ac{BB} \ac{TTD} array where the group delay is introduced in the analog \ac{BB} domain. For each architecture, three types of hardware impairments are considered and they are assumed to be independent across antenna elements and time-invariant.
The frequency flat magnitude mismatch $\alpha_n$ is modeled with a log-normal distribution, i.e., $10\log_{10}(\alpha_n)\sim\mathcal{N}\left(0,\sigma_{\text{A}}^2\right)$. The frequency flat phase error is modeled as $\tilde{\phi}_n\sim\mathcal{N}\left(\phi_n,\sigma_{\text{P}}^2\right)$. The TTD delay error is modeled as $\tilde{\tau}_n\sim\mathcal{N}\left(\tau_n,\sigma_{\text{T}}^2\right)$. Due to HI, the \ac{AWV} in (\ref{eq:TTD_AWV_wo_error}) becomes
\begin{align}
\begin{split}
    \left[ \mathbf{w}_{\text{RF}}[m] \right]_n = &\alpha_n \mathrm{exp}\left[{-j\left(2\pi f_m \tilde{\tau}_n  + \tilde{\phi}_n \right)}\right]\\
    \left[ \mathbf{w}_{\text{BB}}[m] \right]_n = &\alpha_n \mathrm{exp}\left[{-j\left(2\pi (f_m-\fc)\tilde{\tau}_n + \tilde{\phi}_n \right)}\right]
\end{split}
    \label{eq:TTD_AWV_with_error}
\end{align}
for \ac{RF} \ac{TTD} array and \ac{BB} \ac{TTD} array, respectively. 

\textit{Remark 1:} Without \ac{HI}, i.e., $\sigma^2_{\text{A}} = \sigma^2_{\text{P}} = \sigma^2_{\text{T}} = 0$, AWVs in (\ref{eq:TTD_AWV_with_error}) are the same except for the frequency flat phase term $2\pi\fc\tau_n$, which can easily be compensated in $\phi_n$. We note that the two \ac{TTD} architectures are equivalent without \ac{HI}, thus we only discuss the \ac{RF} \ac{TTD} in Sections~\ref{sec:awv_design} and \ref{sec:algorithm}. \ac{HI} will be considered in in \Cref{sec:results}.

\subsection{Problem statement}
In this work, we have two objectives. Firstly, we design the TTD AWV parameters $\tau_n$, $\phi_n$ and $\mathcal{M}$ in (\ref{eq:TTD_AWV_wo_error}) and a beam training algorithm that utilize a single \ac{OFDM} training symbol (\ref{eq:received_signal_sc}) to estimate $\theta^{(\R)}$. Our goal is to improve \ac{AoA} estimation accuracy and robustness in frequency-selective channels, compared to \cite{Yan:TTD}. Secondly, we numerically study the impact of \ac{TTD} hardware impairments on the proposed algorithm with RF \ac{TTD} and BB \ac{TTD} architectures, and discuss feasibility of parameter specifications for hardware implementation.

%
%
\section{\ac{TTD} \ac{AWV} design with frequency diversity}
\label{sec:awv_design}


In this section, we design phase shift and delay taps for the single-shot beam training that incorporates frequency diversity. In \cite{Yan:TTD}, we showed that with uniformly spaced delay and phase taps, i.e., $\tau_n = (n-1)\Delta\tau$ and $\phi_n = (n-1)\Delta\phi$, and properly designed $\Delta\tau$ and $\Delta\phi$, $D$ \ac{DFT} beams can be synthesized simultaneously by $M=D$ selected subcarriers. The $d$-th selected subcarrier has the \ac{AWV} $\mathbf{f}_{d}\in\mathbb{C}^{N_{\R}}$ with the $n$-th element given as
\begin{align}
    [\mathbf{f}_{d}]_n = \mathrm{exp}[-j2\pi (n-1)(d-1-D/2)/D], d\leq D.
    \label{eq:DFT_beams}
\end{align}

However, the one-to-one mapping between frequency and sounding direction in \cite{Yan:TTD} may not be robust in frequency-selective fading. Here, we enhance the design by introducing frequency diversity for \ac{TTD}-based beam training. Intuitively, we divide $\BW$ into $R$ distinct sub-bands. Within each sub-band, we intend to associate $D$ uniformly spaced subcarriers with $D$ fixed \ac{AWV}s $\{\mathbf{f}_d\}_{d=1}^{D}$. Assuming $M_{\tot}/R \in \mathbb{Z}$, the set of indices $\mathcal{M}_d$ of $R$ subcarriers mapped into direction $d$ is 
\begin{align}
    \begin{split}
    \mathcal{M}_d = \Big\{ m ~|~ m =& 1 + (d-1)\lfloor M_{\tot}/(DR)\rfloor\\
    &+(r-1)M_{\tot}/R,~r=1,\cdots,R \Big\}
    \end{split}
    \label{eq:selected_sc}
\end{align}
The set of all subcarriers used in beam training is $\mathcal{M} = \bigcup_{1\leq d\leq D}\mathcal{M}_d$
with cardinality $M =  DR$.
Then, we design $\tau_n$ and $\phi_n$ such that subcarriers in set $\mathcal{M}_d$ have the same \ac{AWV} equal to $\mathbf{f}_d$ (\ref{eq:DFT_beams}), i.e., 
\begin{align}
    \mathbf{w}[m]=\mathbf{f}_d,~m\in\mathcal{M}_d.
    \label{eq:codebook_goal}
\end{align}
As a result, the entire angular range $[-\pi/2, \pi/2]$ is probed by $D$ \ac{DFT} beams simultaneously. The solution to (\ref{eq:codebook_goal}) is provided in the following 
proposition.

%
%
\textit{Proposition 1: }
A feasible way to achieve (\ref{eq:codebook_goal}) is to use uniformly spaced delay and phase taps such that
\begin{align}
    \tau_n = (n-1)R/\BW,\quad \phi_n = (n-1)[\mathrm{sgn}(\psi)\pi - \psi],
    \label{eq:delta_tau_phi}
\end{align}
where $\psi = \mathrm{mod}(2\pi R (\fc-\BW/2)/\BW + \pi, 2\pi) - \pi$. $\mathrm{sgn}()$ and $\mathrm{mod}()$ are the sign and modulo operators, respectively.

\begin{IEEEproof}
This can be simply verified by plugging (\ref{eq:delta_tau_phi}) into (\ref{eq:TTD_AWV_wo_error}) and verifying (\ref{eq:codebook_goal}) for any $d$.
\end{IEEEproof}

%
%
\section{Power based super-resolution beam training}
\label{sec:algorithm}

In this section, we describe the proposed high-resolution \ac{TTD}-based beam training algorithm.
Unlike in \cite{Yan:TTD}, where the dominant \ac{AoA} is simply chosen based on the direction with the highest received signal power, the proposed algorithm in this work jointly considers the received signal powers from all directions to improve the accuracy of \ac{AoA} estimation. 

The frequency selective channel in (\ref{eq:channel1}) can be interpreted as one out of $K_c$ independent realizations of
\begin{equation}
\label{eq:channel2}
    \mathbf{H} = \sum_{l=1}^L G_l \mathbf{a}_{\R}(\theta_l^{(\R)})\mathbf{a}_{\T}^{\hermitian}(\theta_l^{(\T)}),
\end{equation}
where $G_l\sim\mathcal{CN}\left(0, \sigma_{G_l}^2 \right), ~\forall l$. When the diversity order is $R \leq K_c$, $R$ subcarriers from $\mathcal{M}_d$ experience different, independent realizations of the channel in (\ref{eq:channel2}). Consequently, the received signal $Y[m],~m\in\mathcal{M}_d$, in (\ref{eq:received_signal_sc}) is considered as one independent frequency-domain realization of the received signal in direction $d$, given as
\begin{equation}
\label{eq:received_signal_direction}
    Y_d = M^{-1/2}\mathbf{f}_d^{\hermitian} \mathbf{H} \mathbf{v} + \mathbf{f}_d^{\hermitian}\mathbf{n}.
\end{equation}

Since the received signal in (\ref{eq:received_signal_direction}) is a zero-mean complex Gaussian random variable, then  $\abs{Y_d}^2,~\forall d$, is exponentially distributed. 
Based on this model, the expected received signal power in direction $d$ as $p_d = \mathbb{E}\left[ \abs{Y_d}^2 \right] = \mathbb{E}[ (\mathbf{f}_d^{\hermitian} \mathbf{H} \mathbf{v}M^{-1/2} + \mathbf{f}_d^{\hermitian} \mathbf{n} )^{*} (\mathbf{f}_d^{\hermitian} \mathbf{H} \mathbf{v}M^{-1/2} + \mathbf{f}_d^{\hermitian} \mathbf{n} ) ]$.
Due to (\ref{eq:covariance}) and the independence of cluster gains and thermal noise, it follows that  $\mathbb{E}\left[G_{l_1}^* G_{l_2} \right]=0,~l_1\neq l_2$, $\mathbb{E}[ \mathbf{n} G_{l}^* ]=\mathbf{0},~\forall l$, and $\mathbb{E}[G_l \mathbf{n}^{\hermitian} ]=\mathbf{0}^{\transpose},~\forall l$. Since $\mathbb{E}[\abs{G_l}^2]=\sigma_l^2,~\forall l$, the expected received power in direction $d$ can be expressed as
\begin{align}
\begin{split}
    p_d &= \frac{1}{M}\sum_{l=1}^{L} \abs{ \mathbf{f}_d^{\hermitian} \mathbf{a}_{\R}(\theta^{(\R)}_{l}) }^2 \abs{ \mathbf{a}_{\T}^{\hermitian}(\theta^{(\T)}_{l}) \mathbf{v} }^2 \sigma_l^2  + N_{\text{R}}\sigma_N^2\\
    &\approx \frac{1}{M}\abs{ \mathbf{f}_d^{\hermitian} \mathbf{a}_{\R}(\theta^{(\R)}_{1}) }^2 \abs{ \mathbf{a}_{\T}^{\hermitian}(\theta^{(\T)}_{1}) \mathbf{v} }^2 \sigma_1^2  + N_{\text{R}}\sigma_N^2.
\end{split}
    \label{eq:expected_power}
\end{align}
The approximation in (\ref{eq:expected_power}) is due to the fixed precoder $\mathbf{v}$ and large beamforming gain at the BS that result in the received signal with only one spatially filtered dominant cluster. By vectorizing (\ref{eq:expected_power}), we obtain
\begin{equation}
    \mathbf{p} = \mathbf{B} \mathbf{g} + N_{\text{R}}\sigmaN^2\mathbf{1},
    \label{eq:expected_power_matrix}
\end{equation}
where $\mathbf{p} = [ p_1, \cdots, p_D]^{\transpose}$, $\mathbf{B}\in\mathbb{R}^{D\times Q}$, and $\mathbf{g}\in\mathbb{R}^{Q}$. The matrix $\mathbf{B}$ represents a known dictionary obtained by generalizing the receive beamforming gains $\abs{ \mathbf{f}_d^{\hermitian} \mathbf{a}_{\R}(\theta^{(\R)}_{l}) }^2,~\forall l$, using a grid of $Q$ uniformly spaced angles $\xi_q,~q=1,\cdots,Q$. Thus, the $(d,q)$-th element of $\mathbf{B}$ is
\begin{align}
    \left[\mathbf{B}\right]_{d,q} = \left|\mathbf{f}_d^{\hermitian} \mathbf{a}_{\R}(\xi_q)\right|^2.
    \label{eq:dictionary}
\end{align}

The vector $\mathbf{g}$ is a vector of length $Q$, with only one significant element equal to $M^{-1}\abs{ \mathbf{a}_{\T}^{\hermitian}(\theta^{(\T)}_{1}) \mathbf{v} }^2 \sigma_1^2$, corresponding to the dominant cluster.

According to discussion in \Cref{sec:awv_design}, $\abs{Y_d}^2$ has $R$ independent realizations $\abs{Y[m]}^2,~m\in\mathcal{M}_d$. We propose to use the maximum likelihood estimator to evaluate $p_d$:
\begin{equation}
    \hat{p}_d = \frac{1}{R} \sum_{m\in\mathcal{M}_d} \abs{Y[m]}^2.
    \label{eq:sample_mean}
\end{equation}
The estimate $\hat{\mathbf{p}}$ is obtained by vectorization $\hat{\mathbf{p}}=\left[\hat{p}_1, \cdots, \hat{p}_D \right]^{\transpose}$. Based on (\ref{eq:expected_power_matrix}), the problem of estimating $\theta^{(\R)}$ reduces to the problem of finding the index of the element $M^{-1}\abs{ \mathbf{a}_{\T}^{\hermitian}(\theta^{(\T)}_{1}) \mathbf{v} }^2 \sigma_1^2$ in $\mathbf{g}$. Equivalently, one can find the index of the column of $\mathbf{B}$ which has the highest correlation with $\hat{\mathbf{p}}$. The proposed \ac{AoA} estimate is
\begin{equation}
    \hat{\theta}^{(\R)} = \xi_{q^{\star}}, \text{ where } q^{\star} = \argmax_q \frac{\hat{\mathbf{p}}^{\transpose}[\mathbf{B}]_{:,q}}{||[\mathbf{B}]_{:,q}||}.
    \label{eq:angle_estimate1}
\end{equation}

The proposed beam training scheme and algorithm are summarized in \textbf{Algorithm 1}. The resolution of the algorithm is determined by $Q$ and its complexity is $\mathcal{O}(M+DQ)$.

\begin{algorithm}
\caption{TTD array based super-resolution beam training}
\begin{algorithmic}[1]
\Statex \textbf{Input:} \ac{UE} analog array settings in (\ref{eq:delta_tau_phi}). Pre-computed dictionary in (\ref{eq:dictionary}). A single received OFDM symbol $Y[m], m\in \mathcal{M}$, with subcarrier selection in (\ref{eq:selected_sc}).
\Statex \textbf{Output:} \ac{AoA} estimate $\hat{\theta}^{(\R)}$.
  \State Compute direction powers based on (\ref{eq:sample_mean})
  \State Use (\ref{eq:angle_estimate1}) to find AoA estimate $\hat{\theta}^{(\R)}$
\end{algorithmic}
\end{algorithm}

\section{Performance Results}
\label{sec:results}
In this section, we numerically study the impact of diversity order $R$ and \ac{HI} on the proposed super-resolution beam training algorithm. The algorithm from \cite{Yan:TTD} is extended with proposed frequency diversity scheme and used as the benchmark. We consider a system with carrier frequency $\fc=$ \SI{60}{\giga\hertz}, bandwidth $\BW=$ \SI{2}{\giga\hertz}, and $M_{\tot}=4096$ subcarriers.
The transmitter and receiver array size are $N_{\T}=128$ and $N_{\R}=16$, while the resolution of the \ac{ADC} is set to 5 bits. The number of sounded directions in beam training is $D=32$ and the dictionary size is $Q=1024$. The channel consists of $L=3$ clusters (1 strong, 2 weak, with $10$dB relative difference). Fading is simulated by 20 rays within each cluster with \SI{10}{\nano\second} spread, i.e., $K_c \leq 20$. There is no intra-cluster angular spread. The \ac{SNR} is defined as $\SNR \triangleq \sum_{l=1}^{L}  \sigma_l^2/\sigmaN^2$.

\begin{figure}
    \begin{center}
        \includegraphics[width=0.48\textwidth]{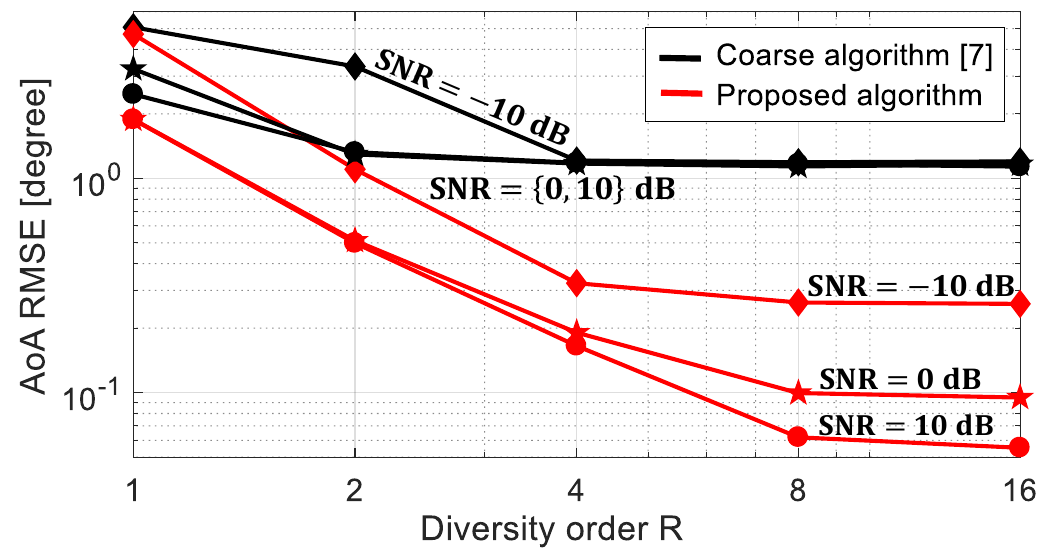}
    \end{center}
    \vspace{-4mm}
    \caption{The impact of diversity order on the algorithms.}
    \vspace{-4mm}
    \label{fig:diversity}
\end{figure}

\Cref{fig:diversity} presents the impact of diversity order $R$ for different values of \ac{SNR}. The coarse-resolution algorithm improves with diversity for $R\leq4$ , but experiences an error floor due to limited resolution. On the other hand, the proposed algorithm's AoA RMSE is significantly lower and monotonically decreases with $R$ for any \ac{SNR} until it reaches the error floor. The resolution of the proposed algorithm is limited by dictionary size $Q$, i.e., accuracy up to $\pm \pi/(2Q)$ and thus $\mathrm{RMSE}\geq \sqrt{(\pi/Q)^2/12}$ in a high \ac{SNR} regime.

\begin{figure}
    \begin{center}
        \includegraphics[width=0.48\textwidth]{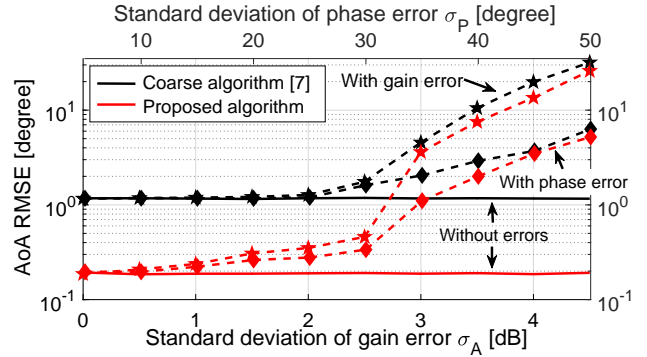}
    \end{center}
    \vspace{-4mm}
    \caption{The impact of gain error and phase error on the algorithms. The curves with gain error (dashed with stars) and phase error (dashed with diamonds) are associated with the lower and upper x-axis, respectively.}
    \vspace{-3mm}
    \label{fig:gain_phase}
\end{figure}

Next, we study the impact of hardware impairments, assuming $\SNR= $ \SI{0}{\decibel}. In the evaluation, the received signal (\ref{eq:received_signal_sc}) is generated using $\mathbf{w}_{\text{RF}}[m]$ or $\mathbf{w}_{\text{BB}}[m]$, while the algorithm's dictionary in (\ref{eq:dictionary}) is based on impairment-free \ac{AWV} $\mathbf{w}[m]$. The diversity order is $R=4$. In \Cref{fig:gain_phase}, we study the impact of gain and phase errors, under no delay error. Note that $\mathbf{w}_{\text{RF}}[m] = \mathbf{w}_{\text{BB}}[m]$ which implies both architectures behave the same with these types of impairments. For both algorithms, severe performance degradation occurs when gain error $\sigma_{\text{A}}\geq$ \SI{2.5}{\decibel} and phase error $\sigma_{\text{P}}\geq 0.52$ (\SI{30}{\degree}). 

\begin{figure}
    \begin{center}
        \includegraphics[width=0.48\textwidth]{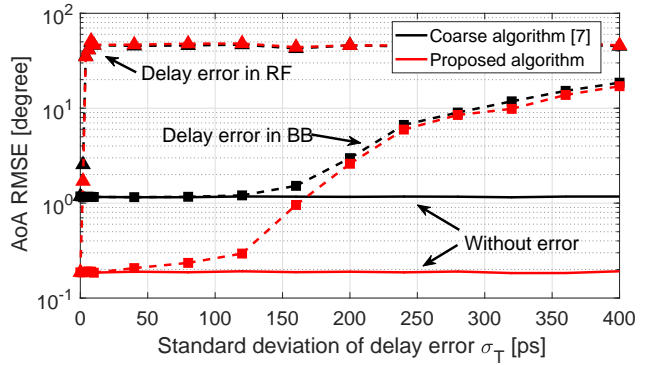}
    \end{center}
    \vspace{-4mm}
    \caption{The impact of TTD delay error in both array architectures on the algorithms.}
    \vspace{-6mm}
    \label{fig:delay}
\end{figure}

In \Cref{fig:delay}, the impact of \ac{TTD} delay error is presented. Using \ac{BB} TDD architecture, both algorithms are robust to delay errors with standard deviation of up to $\sigma_{\T}=$ \SI{125}{\pico\second}. In comparison, both algorithms start to show severe degradation with only $\sigma_{\T}=$\SI{1.5}{\pico\second} in \ac{RF} TDD architectures. This result indicates that both algorithms have more relaxed specifications for BB vs. RF implementation of \ac{TTD} elements.

Besides the three studied impairments, certain hardware constraints can also affect the performance of the beam training algorithms, including finite \ac{TTD} delay tap resolution and maximum \ac{TTD} delay range. Beam training without diversity ($R=1$) requires the finest resolution since 
$\Delta\tau = 1/\BW =$
 \SI{0.5}{\nano\second} \cite{Yan:TTD}. As we consider the case with $R=4$, the resolution requirement can be relaxed to $\Delta\tau = 4/\BW =$ \SI{2}{\nano\second}. On the other hand, larger diversity order $R$ imposes larger requirement on the maximum delay range $\tau_{N_{\R}}=(N_{\R}-1)R/\BW$. In the considered case with $N_{\R}=16$ and $R=4$, the maximum delay range should be at least $\tau_{N_{\R}}=$ \SI{30}{\nano\second}. 


%
%
\vspace{0mm}
\section{Efficient implementation of TTD array}
\label{sec:implementation}


This section briefly describes circuits design considerations and implementation of the TTD arrays for the proposed \ac{mmW} beam training algorithm. Conventionally, implementation of delay compensating elements at RF has been adopted in several approaches. However, large delay range and resolution necessitates rethinking of the array design as the range-to-resolution ratios in RF TTD arrays have been limited due to constraints from linearity, noise, area, and tunability. Additionally, our results in \Cref{sec:results} showed that \ac{RF} \ac{TTD} arrays are sensitive to even small delay errors.

\begin{figure}[t]
\begin{tabular}{cc}
\vspace{-4mm}
\subfloat[]{%
  \includegraphics[clip,width=0.45\textwidth]{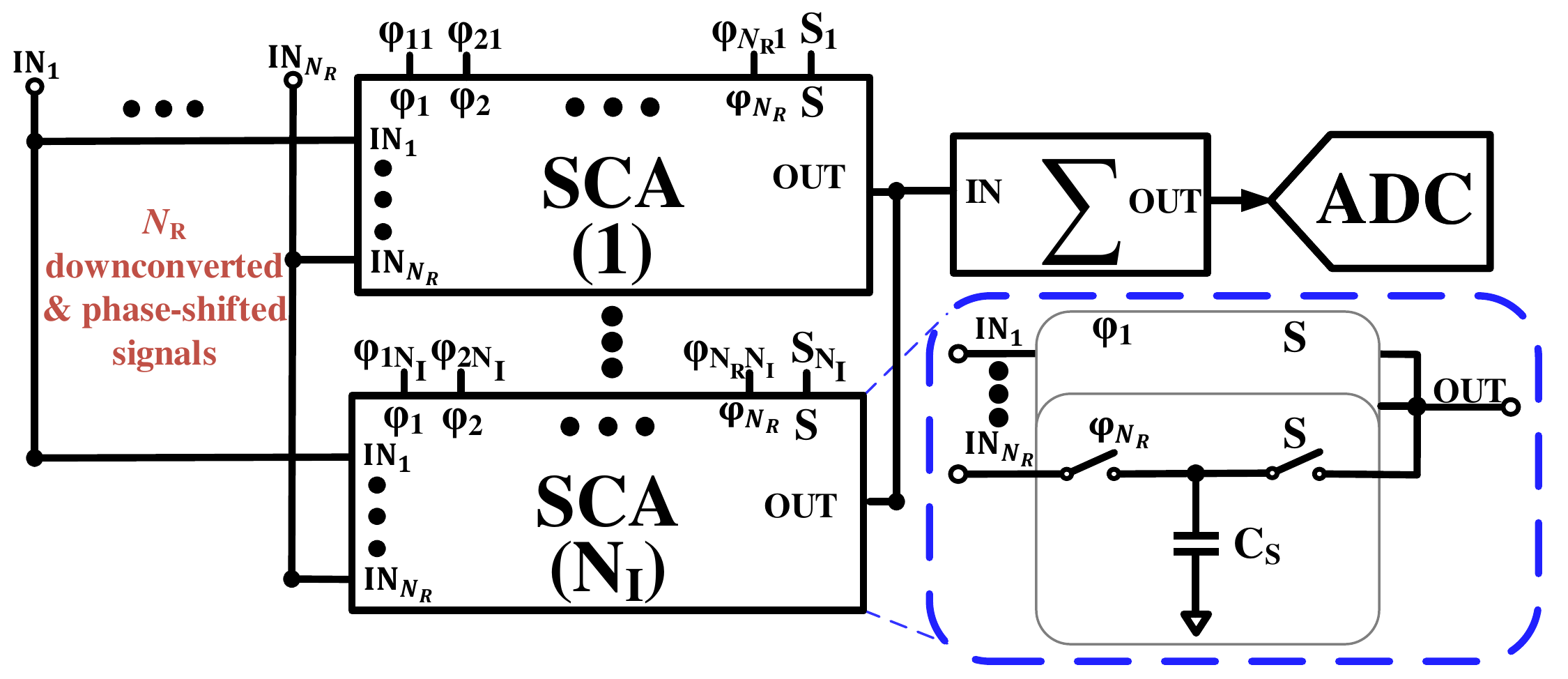}%
}\\
\vspace{0mm}
\subfloat[]{%
  \includegraphics[clip,width=0.45\textwidth]{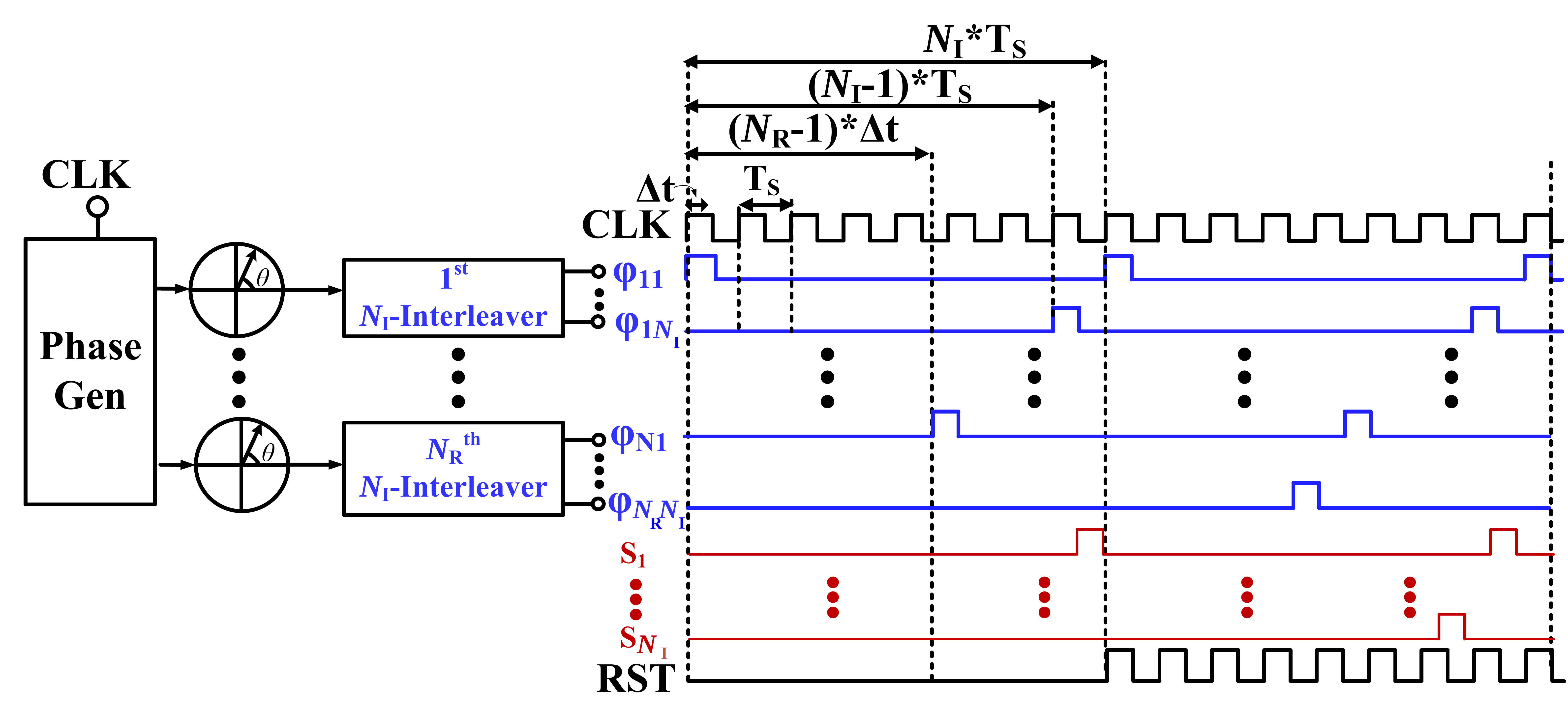}%
}\\
\vspace{-4mm}
\end{tabular}
\caption{(a) TTD module based on analog SCA and (b) time-interleaving circuit for large delay range and resolution \cite{ghaderi2019}.}
\vspace{-6mm}
\label{fig:WSU1}
\end{figure}

In \cite{ghaderi2019}, we demonstrated a discrete-time delay compensation technique that enabled realization of large range-to-resolution ratios using a baseband \ac{TTD} element. In this technique, instead of delaying the down-converted and phase-shifted signals from $N_{\R}$ antennas and then sampling and digitizing, they are sampled at different time instants through the \ac{SCA}, resulting in the same digitized value as shown in \Cref{fig:WSU1}(a). Thus, the complexity of delaying in signal path at high \ac{RF} or even down-converted frequency, i.e., analog baseband, is shifted to the clock path where precise and calibrated delays can be applied in nanometer \ac{CMOS} technologies. More importantly, this enables large delay range-to-resolution ratios to be realized. For example, in \cite{ghaderi2019}, we can achieve up to \SI{15}{\nano\second} delay range with \SI{5}{\pico\second} resolution supporting \SI{100}{\mega\hertz} signal bandwidth. This bandwidth was recently enhanced to \SI{500}{\mega\hertz} in \cite{ghaderi2020} leveraging the digital-friendly and technology-scalable time-based circuits. The \ac{SCA} based implementation of the beamformer requires $N_{\R}$ phases for sampling ($\left\{\varphi_n\right\}_{n=1}^{N_{\R}}$) followed by addition (S) and \ac{RST} phase, in an $N_{\R}$-element array. In the sampling phase, $\left\{\varphi_n\right\}_{n=1}^{N_{\R}}$, an input signal from each antenna receiver is first sampled on a sampling capacitor $C_S$ uniquely. After the last sampling phase ($\varphi_{N_{\R}}$), the stored charges on each capacitor are shared in the S phase. This charge sharing performs an averaging function. To change this functionality to summation and form the beam, the shared charges are transferred to the feedback capacitor in a switched-capacitor summer following the \ac{SCA}. After forming the beam, the summer is reset by the \ac{RST} phase to prepare for the next sample. For proper functionality, S should be non-overlapping with all the sampling phases ($\left\{\varphi_n\right\}_{n=1}^{N_{\R}}$) and the \ac{RST} phase. To compensate the time delay between the $N_{\R}$ inputs in this \ac{SCA}, each signal must be sampled at a distinct time instant. Hence, both the sampling ($\left\{\varphi_n\right\}_{n=1}^{N_{\R}}$) and addition phases (S) are further time-interleaved to $N_{\I}$ phases as shown in \Cref{fig:WSU1}(b). The period of each time-interleaved phase is set to $N_{\I}$ times of the initial phase period.


Larger diversity orders however create further challenges to the analog \ac{TTD} array implementation with higher number of interleaving levels that are required. Table I highlights the number of interleaving levels required for different diversity orders for a $N_{\R}=16$ element array with \SI{2}{\giga\hertz} bandwidth. Our current TDD analog array implementation is suitable for smaller diversity orders. Larger delay ranges required for higher diversity orders can be implemented more efficiently using  hybrid (analog/digital) TTD array architectures \cite{ghaderi2019}. 

\begin{table}[h]
\vspace{-2mm}
\centering
\caption{Analog TTD array complexity with increased diversity $R$.}
\label{tab:T1}
\vspace{-2mm}
{   
    \small
    \begin{tabular}{|c||c||c||c||c||c|} 
    \hline
        \multirow{2}{*}{$R$}  & \multirow{2}{*}{$R/\BW$} & $\tau_{N_{\R}}$ & $N_{\I}$ &  $\tau_{N_{\R}}$ & $N_{\I}$\\ & & ANA & ANA  & ANA-DIG & ANA-DIG\\
        \hline
        \hline
        1  &  \SI{0.5}{\nano\second} & \SI{7.5}{\nano\second} & 31 & - & -  \\
        \hline
        2  &  \SI{1}{\nano\second} & \SI{15}{\nano\second} & 61 & \SI{2}{\nano\second} & 9        \\
        \hline 
        4  &  \SI{2}{\nano\second} & \SI{30}{\nano\second} & 121 & \SI{6}{\nano\second} & 25 \\
        \hline 
    \end{tabular} \\
}
\vspace{1mm}
ANA: analog TTD array; ANA-DIG: hybrid TTD array with four 4-element sub-arrays; 
$N_{\R} = 16, f_{\text{CLK}} =$ \SI{4}{\giga\hertz}, $\BW = $ \SI{2}{\giga\hertz}. 
\end{table}
\vspace{-2mm}


%
%
\section{Conclusions}
\label{sec:conclusions}
In this work, we developed TTD based single-shot super-resolution beam training that exploits frequency diversity. The proposed algorithm is robust to frequency selective fading and hardware impairments when the TTD array is implemented in baseband. Based on this finding, we provided an insight on feasible implementation of a TTD array.

%
%

\section{Acknowledgement}
This work was supported in part by NSF under grants 1718742 and 1705026. This work was also supported in part by the ComSenTer and CONIX Research Centers, two of six centers in JUMP, a Semiconductor Research Corporation (SRC) program sponsored by DARPA.

\bibliographystyle{IEEEtran}
\bibliography{IEEEabrv,references}

\end{document}